\documentstyle[twocolumn,prl,aps,amssymb,epsfig]{revtex}
\begin{document}
\twocolumn[\hsize\textwidth\columnwidth\hsize\csname@twocolumnfalse\endcsname
\draft

\title{Dynamical Scaling Properties of Electrons in Quantum Systems
with \\Multifractal Eigenstates}

\author{Jianxin Zhong,$^{1,2,3}$ Zhenyu Zhang,$^{2,1}$ Michael
        Schreiber$^{4}$, E. Ward Plummer,$^{1,2}$ and Qian
        Niu$^{2,5}$}

\address{ $^1$Department of Physics, University of Tennessee,
Knoxville, Tennessee 37996\\ $^2$Solid State Division, Oak Ridge
National Laboratory, Oak Ridge, Tennessee 37831\\
$^3$Department of Physics, Xiangtan University, Hunan 411105, P.R. China\\
$^4$Institut f\"{u}r Physik,
Technische Universit\"{a}t, D-09107 Chemnitz, Germany\\ $^5$Department
of Physics, University of Texas at Austin, Austin, Texas 78712}


\maketitle

\begin{abstract}%

 We study the intricate relationships between the dynamical scaling
 properties of electron wave packets and the multifractality of
 the eigenstates in quantum systems.  Numerical simulations for the Harper
 model and the Fibonacci chain indicate that the root mean square
 displacement displays the scaling behavior $r(t)\sim t^\beta$ with
 $\beta=D_2^\psi$, where $D_2^\psi$ is the correlation dimension of
 the multifractal eigenstates. The equality can be generalized to
 $d$-dimensional systems as $\beta=D_2^\psi/d$, as long as the
 electron motion is ballistic in the effective $D_2^\psi$-dimensional
 space. This equality should be replaced by $\beta<D_2^\psi/d$ if the
 motion is non-ballistic, as supported by all known results.
\end{abstract}

\pacs{PACS numbers:
 71.23.Ft,  
 71.30.+h,  
 72.15.Rn   
 05.45.+b,  
 }
]

\narrowtext

Spatial extension of eigenstates influences electron dynamics and
transport properties of solids. It is well known that extended
eigenstates in periodic lattices result in ballistic motion of
electrons while localized eigenstates in strongly disordered systems
lead to zero mobility. As a consequence, one expects metallic phase
and insulating phase, respectively. In between these two extremes
stand the critical eigenstates, i.e., the multifractal eigenstates,
which are neither extended nor exponentially localized. Multifractal
eigenstates exist in various quantum model systems, such as the Harper
model of two-dimensional (2D) electrons in the presence of a
perpendicular magnetic field\cite{HP}, tight-binding models of
quasicrystals \cite{MBC,FT,FIB,ZGRS}, and disordered systems at the
metal-insulator transition (MIT) \cite{SE,FG,SG,BHS,OK,TT,HK,SKO}. In
recent years, a great deal of effort has been devoted to the analysis
of the multifractal nature of the eigenstates. Effort has also been
made to establish intrinsic connections between the multifractality of
the eigenstates and the transport properties in such systems. As an
example, the unusual resistivity increase caused by decreasing
temprature or improving structural quality of quasicrystals has been
attributed to the multifractal nature of the electronic states
\cite{MBC}.  Despite all these efforts, a fundamental level of
understanding on precisely how the multifractality of the eigenstates
affects the electron dynamics is still lacking.
                           
The characteristics of electron dynamics is reflected by the time
evolution of the corresponding wave packet, whose spreading can be
described by the root mean square displacement $r(t)$. For ballistic
spreading, $r(t)\sim t$, while in the absence of diffusion, $r(t)\sim
t^0$ for large $t$. Numerical simulations \cite{AH} for the Harper
model and the Fibonacci quasiperiodic chain \cite{HP,MBC,FT,FIB}
showed that $r(t)\sim t^\beta$ with $0<\beta<1$. Such a scaling
behavior has also been found in other systems with multifractal
eigenstates \cite {OK,HK,UT,PSB}. On the other hand, the energy
spectrum of the multifractal eigenstates can be measured by the
multifractal dimensions, $D_q^\mu$ \cite{HJKPS}. The second moment,
$D_2^\mu$, has been shown to determine the decay of the average
probability, $C(t)$, of a wave packet remaining at its initial
position: $C(t)\sim t^{-D_2^\mu}$ \cite{KPG,ZMB}. It is naturally
tempting to link the spectral multifractal dimensions, $D_q^\mu$, with
the dynamical scaling exponent, $\beta$
\cite{G,GKGP,EKWA,M,P,KKKG,BB}. Indeed, earlier such efforts have led
to a variety of results, including $\beta\leq D_1^\mu$ \cite{G},
$\beta =D_{-1}^\mu$ \cite{M,P}, and $\beta\geq D_2^\mu/D_2^\psi$
\cite{KKKG}, where $D_2^\psi$ is the correlation dimension of the
eigenstates.  We note that all these results were derived with the
assumption that the energy spectra are singularly continuous. Scaling
behavior of $r(t)$ may also exist in higher dimensional systems whose
spectra are not singularly continuous \cite{UT,PSB,ZMB}, but it would
be even more difficult to derive a unique relation between $\beta$ and
$D_q^\mu$.

In this Letter, we use a new approach to understand the scaling
behavior of electron dynamics.  Instead of focusing on the {\sl
energy} spectra, we study how the multifractality of the eigenstates
in {\sl real} space affects the time spreading of the wave
packets. Our study leads to several fundamentally important
findings. First, for the 1D Harper model and the Fibonacci chain, our
numerical simulations show that $\beta$ is directly given by the
correlation dimension of the multifractal eigenstates:
$\beta=D_2^\psi$. Second, by interpreting $D_2^\psi$ as the effective
dimension of the multifractal space for electron motion, we generalize
the above relation to $d$-dimensional systems as
$\beta=D_2^\psi/d$. This equality holds as long as the spreading of
wave packets in the $D_2^\psi$-space is ballistic in nature, as
demonstrated by applying to exactly solvable 2D and 3D quasiperiodic
systems. Furthermore, when the spreading in $D_2^\psi$-space is
non-ballistic, we have $\beta<D_2^\psi/d$, which is in complete
agreement with all the known results for higher dimensional ($d>1$)
disordered systems \cite {BHS,OK,TT,HK,SKO}.

We start with a 1D lattice of size $L=Na$, where $N$ is the total
number of sites and $a$ the lattice constant. The Hamiltonian has the
tight-binding form $H=\sum_n\epsilon_n|n><n|+\sum_{m,n}t_{m,n}|m><n|$,
where $\epsilon_n$ is the site energy, $t_{m,n}$ the nearest-neighbor
hopping elements.  Eigen-energies and eigenstates are denoted as $E_m$
and $\psi_n(E_m)$, respectively. The multifractal analysis of an
eigenstate is based on the standard box-counting procedure: divide the
system into a number of boxes of linear size $wL$ $(0<w<1)$, and then
calculate the box probability in the $k$th box, $p_k=\sum_{n\in
k}|\psi_n(E_m)|^2$.  A spatial correlation function $R(w,E)$ is
defined as $R(w,E)=\sum_k p_k^2$. In general, $R(w,E)\sim
w^{D_2^\psi(E)}$, where $D_2^\psi(E)$ is called the correlation
dimension of the eigenstate. Extended eigenstates and localized
eigenstates are characterized by $D_2^\psi(E)=1$ and $D_2^\psi(E)=0$,
respectively. For multifractal eigenstates, $0<D_2^\psi(E)<1$.
Considering the fact that eigenstates contributing to the time
spreading of a wave packet may display different multifractal
behaviors, we introduce a spectral-averaged spatial correlation
function, $R(w)={1\over N}\sum_mR(w,E_m)$, and define a
spectral-averaged correlation dimension $D_2^\psi$ if $R(w)\sim
w^{D_2^\psi}$ holds.  The root mean square displacement for a wave
packet initially located at site $n_0$ is given by
$r(t,n_0)=\sqrt{\sum_n|n-n_0|^2|\phi_n(t)|^2}$, where the wave
function $\phi_n(t)$ can be obtained by
$\phi_n(t)=\sum_m\psi^*_{n_0}(E_m)\psi_n(E_m)\exp(iE_mt)$.  For
inhomogeneous systems, $r(t,n_0)$ varies with $n_0$.  In this case,
one averages $r(t,n_0)$ over different initial wave packets in order
to obtain the dynamics for the whole system. The scaling behavior of
the system-averaged root mean square displacement is defined as $r(t)
\sim t^\beta$, with $0<\beta<1$.

\vspace{1cm}
\begin{figure}
  \centerline{\hspace{1cm}\epsfig{figure=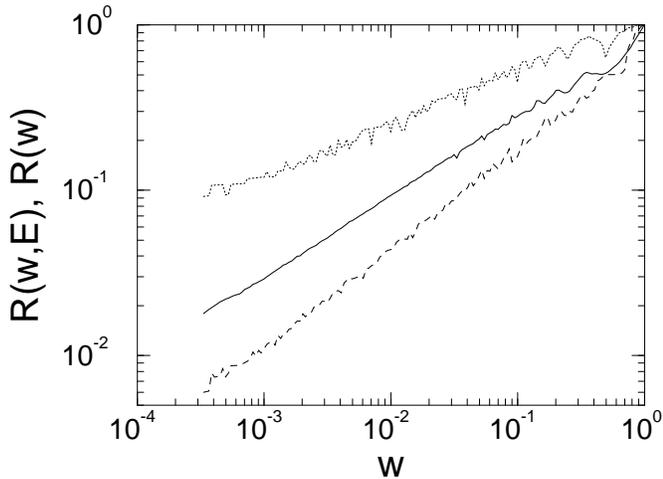,width=0.75\columnwidth}}
   \vspace{1.5cm}
  \caption{Multifractal properties of eigenstates for the Harper model
  of system size $N=3000$. The solid line shows $R(w)\sim
  w^{D_2^\psi}$ with $D_2^\psi=0.5$.  The dotted and the dashed lines
  show $R(w,E)\sim w^{D_2^\psi(E)}$ for energies $E_e$ and $E_c$ with
  $D_2^\psi(E_e)=0.38$ and $D_2^\psi(E_c)=0.64$, respectively.}
\end{figure}

Hamiltonian parameters in the Harper model \cite{HP} are $t_{m,n}=1$
and $\epsilon_n=2\cos(2\pi\sigma n)$, where $\sigma =(\sqrt{5}-1)/2$
is the golden mean. We find that the correlation function $R(w,E)$ for
various eigenstates displays a scaling behavior $R(w,E)\sim
w^{D_2^\psi(E)}$, where $0<D_2^\psi(E)<1$ and $D_2^\psi(E)$ is
different for different eigenstate. As examples, we show $R(w,E)$ for
a system of $N=3000$ in Fig.~1 at energies $E_{\rm e}=2.59751453292$
and $E_{\rm c}=0.00000066284$, which respectively correspond to the
upper edge and the center of the energy spectrum. The corresponding
scaling exponents are $D_2^\psi (E_{\rm e})=0.38$ and $D_2^\psi
(E_{\rm c})=0.64$.  We also find that the spectral-averaged (over the
whole spectrum of $-2.6<E<2.6$) correlation function $R(w)$ displays a
scaling behavior $R(w)\sim w^{D_2^\psi}$ with $D_2^\psi=0.5$ as shown
in Fig.~1. In Fig.~2, $R(w)$ is compared with $r(t)$ obtained by
averaging over $500$ initial wave packets. We find $r(t)\sim t^\beta$
with $\beta=0.5$, in agreement with a previous study \cite{P}. We
therefore have $\beta=D_2^\psi=0.5$.

\vspace{1cm}
\begin{figure}
  \centerline{\hspace{1cm}\epsfig{figure=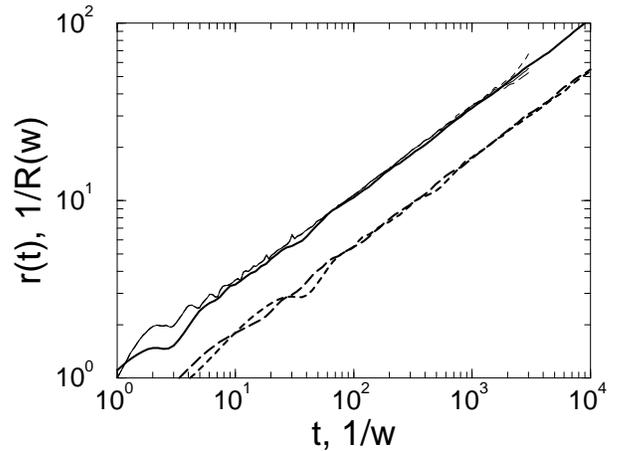,width=0.7\columnwidth}}
  \vspace{1.5cm} 	
  \caption{Scaling behaviors of $r(t)\sim t^\beta$ (thick lines) and
  $R(w)\sim w^{D_2^\psi}$ (thin lines) for the whole spectrum (solid
  lines), the center sub-spectrum (dashed lines), and the edge
  sub-spectrum (long dashed lines) in the Harper model with
  $\beta=D_2^\psi=0.5$. System size is $N=3000$.}
\end{figure}

We can also study how eigenstates in different spectral regions affect
the spreading of wave packets by considering
$\phi_n(t)=\sum_{E_m\in\Omega}\psi^*_{n_0}(E_m)\psi_n(E_m)\exp(iE_mt)$,
where $\Omega$ denotes the spectral region. The corresponding $r(t)$
is compared with $R(w)$ obtained by averaging the eigenstates within
$\Omega$.  Both the spectrum and the eigenstates of the Harper model
have self-similar trifurcating structures
\cite{HP,MBC,FT,FIB}. Therefore, we expect that the spreading of the
wave packets contributed by eigenstates in different sub-spectra
follow the same scaling law. In Fig.~2, we show $R(w)$ and $r(t)$ for
the center sub-spectrum $-0.2<E<0.2$ and the edge sub-spectrum
$1.8<E<2.6$. It is clear that $R(w)$ for different sub-spectra is the
same because of the same multifractal behavior of the eigenstates in
different sub-spectra.  Furthermore, $r(t)$ for different sub-spectra
in Fig.~2 displays the same scaling law with $\beta=D_2^\psi=0.5$, as
expected. This indicates that the multifractality of the eigenstates
indeed plays a crucial role in the spreading of the wave packets. We
note that the downward shift of $r(t)$ shown in Fig.~2 for the
sub-spectra is due to the decrease in the number of eigenstates
contributing to the spreading of the wave packets.

We find the same relation $\beta=D_2^\psi$ to hold for the Fibonacci
quasiperiodic chain \cite{FIB}.  In this model, $t_{i,j}=1$ and
$\epsilon_i$ takes $-u$ and $u$ arranged in a Fibonacci sequence. The
averaged correlation function has the scaling behavior $R(w)\sim
w^{D_2^\psi}$ with $0<D_2^\psi<1$ depending on $u$. The
system-averaged root mean square  displacement $r(t)$ used to be compared
with $R(w)$ was obtained by an average of $500$ wave packets.  As
examples, we show $R(w)$ and $r(t)$ for the whole spectrum in Fig.~3
around $u=1.65$: $\beta=D_2^\psi =0.62$ for $u=1$; $\beta=D_2^\psi
=0.5$ for $U=1.65$; and $\beta=D_2^\psi =0.43$ for $u=2$. These three
cases cover the super-diffusive, diffusive, and sub-diffusive regimes
of motion in real space.

\vspace{1cm}
\begin{figure}
\centering {\hspace{1cm}\epsfig{figure=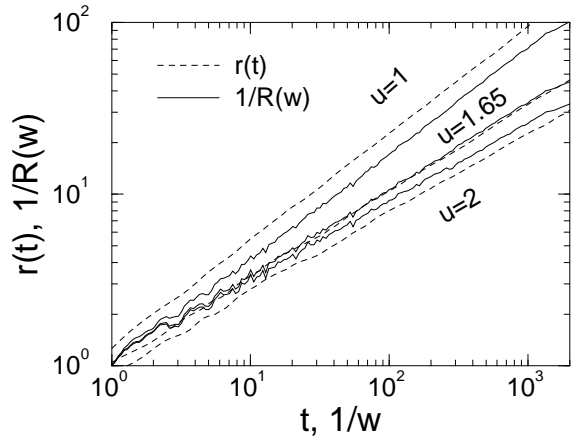,width=0.7\columnwidth}}
  \vspace{1.5cm}
  \caption{Scaling behaviors of $r(t)\sim t^\beta$ (dashed lines) and
  $R(w)\sim w^{D_2^\psi}$ (solid lines) in the Fibonacci chain with
  site energies $u=1, 1.65$, and $2$. The corresponding exponents are
  $\beta=D_2^\psi=0.62, 0.5$, and $0.43$, respectively. System size is
  $N=2584$.}
\end{figure}
 
The relation $\beta=D_2^\psi$ observed in different models can be
generalized to higher dimensional systems with multifractal
eigenstates. From its definition, $R(w)$ is in fact the average
probability for an electron in a box of linear size $wL$.  Because
$R(w)\sim w^{D_2^\psi}$, the effective number of boxes of the system
is $N'=({1\over w})^{D_2^\psi}$, instead of $(1/w)^d$, in order to
ensure the normalization $N'\times R(w)\sim 1$. From the standard
definition of the volume dimension $D_v$, i.e., {\sl the number of the
measured boxes}$\times (wL)^{D_v}=constant$, we have
$D_v=D_2^\psi$. Thus $D_2^\psi$ can be viewed as the effective volume
dimension for electron motion in the multifractal space (hereafter
called the $D_2^\psi$-space). If the motion (i.e., the spreading of
the wave packet) is ballistic in $D_2^\psi$-space, the volume, $N'$,
visited by the electron is $N'\sim t^{D_2^\psi}$. Because the width of
the wave packet $r(t)$ is still measured in the $d$-dimensional real
space, we have $r(t)\sim t^{D_2^\psi/d}$, which immediately leads to
the result $\beta=D_2^\psi/d$.

As tests of the above generalized relation, we consider the exactly
solvable models in higher dimensional Fibonacci lattices \cite{UT,ZMB}
and the Labyrinth lattices \cite{S}.  We find that the equality
$\beta=D_2^\psi/d$ holds for all those models. As an example, here we
only show the case for 3D Fibonacci lattice, which is constructed in a
cubic lattice with constant hopping elements $t_{i,j}=1$ and separable
site energy $\epsilon_i=\epsilon_{ix}+\epsilon_{iy}+\epsilon_{iz}$,
where $\epsilon_{ix}, \epsilon_{iy}$, and $\epsilon_{iz}$ are the 1D
Fibonacci site energies. The energy spectrum is given by
$E=E_x+E_y+E_z$ and the corresponding eigenstates are
$\psi(E,x,y,z)=\psi_x(E_x)\times \psi_y(E_y)\times \psi_z(E_z)$, where
$E_j$ and $\psi_j(E_j)$ $(j=x,y,z)$ are the energy spectra and the
eigenstates for the 1D Fibonacci chain, respectively. As a
consequence, these eigenstates are multifractal \cite{UT,ZMB}, and
their spreading dynamics is non-ballistic in real space: $r(t)\sim
t^\beta$ \cite{UT}, where $0<\beta<1$ takes the same value as that for
the 1D Fibonacci chain.  We note that all the relationships between
the energy spectra and the dynamics \cite{G,GKGP,EKWA,M,P,KKKG} cannot
give correct prediction for the spreading of wave packets in these
higher dimensional quasiperiodic systems. In our approach, for the
higher dimensional Fibonacci lattices, it is easy to show that $R(w)$
has the scaling behavior with $D_2^\psi=dD_2^{\psi_x}$, where
$D_2^{\psi_x}$ is the multifractal dimension of the eigenstates in the
1D Fibonacci chain. Therefore, we have $\beta=D_2^\psi/d$.

Electron motion in the effective $D_2^\psi$-space can also be
non-ballistic. In such cases, the above equality should be replaced by
$\beta<D_2^\psi/d$, because no motion should be faster than ballistic.
One special example is the diffusive motion ($\beta=0.5$) in the
metallic regime of 3D disordered systems with extended eigenstates
($D_2^\psi=3$), where $\beta<D_2^\psi/3$.  As shown below, this
inequality is also satisfied by many other known examples involving
non-ballistic motion in the $D_2^\psi$-space. At the metal-insulator
transition of 3D disordered systems (in the absence of magnetic
field), the eigenstates are found to be multifractal with different
values of $D_2^\psi$: $D_2^\psi/3=0.57\pm 0.1$\cite {SE},
$D_2^\psi/3=0.48-0.6$\cite {SG}, $D_2^\psi/3=0.57\pm 0.07$ \cite
{BHS}, and $D_2^\psi/3=0.5\pm 0.07$ \cite {OK}. For such systems in a
magnetic field, we have $D_2^\psi/3=0.57\pm 0.07$ \cite {OK},
$D_2^\psi/3\approx 0.5$ \cite {TT}, and $D_2^\psi/3=0.43\pm 0.03$
\cite{HK}.  On the other hand, the corresponding electron dynamics was
found to have the same scaling exponent $\beta=0.33$ \cite {OK},
showing that $\beta<D_2^\psi/d$ for all these cases (with or without a
magnetic field). Existing numerical results for 2D disordered systems
satisfy this inequality as well. For the 2D disordered systems in a
strong magetic field, eigenstates at the centers of the Landau bands
are multifractal with $D_2^\psi/2=0.76\pm 0.03$ and the corresponding
spreading has a scaling behavior with $\beta=0.5$ \cite{HK}. At the
MIT of the 2D disordered systems with spin-orbit interactions, one
finds $D_2^\psi/2=0.84\pm 0.03$ and $\beta=0.5$ \cite{SKO}. It is
evident that $\beta< D_2^\psi/d$.
              
In conclusion, we have shown that there exists an intrinsic
relationship between the dynamic spreading of electron wave packets
and the multifractality of the eigenstates in quantum systems. For the
1D Harper model and the Fibonacci chain, we have $\beta=D_2^\psi$,
where $\beta$ is the scaling exponent of the root mean square
displacement, and $D_2^\psi$ the correlation dimension of the
multifractal eigenstates. We have further provided a conceptually new
interpretation of $D_2^\psi$ as the effective dimension for electron
motion in the multifractal space. Following this interpretation, the
above equality can be generalized to $d$-dimensional systems as
$\beta=D_2^\psi/d$, as long as the electron motion is ballistic in the
$D_2^\psi$-space. We have demostrated that such an equality applies to
exactly solvable 2D and 3D quasiperiodic systems.  When the motion in
the $D_2^\psi$-space is non-ballistic due to disorder, the equality is
replaced by $\beta<D_2^\psi/d$, which is in complete agreement with
all the known results for higher dimensional ($d>1$) disordered
systems.

We thank T. Geisel, R. Ketzmerick, J. Bellissard, R.A. R{\"o}mmer,
U. Grimm, X.C. Xie and G. Canright for helpful discussions. This
research was supported by the US National Science Foundation under
Grant No. DMR-9702938, by Oak Ridge National Laboratory, managed by
Lockheed Martin Energy Research Corp. for the Department of Energy
under contract No. DE-AC05-96OR22464, and in part by the National
Natural Science Foundation of China.

\end{document}